\documentclass[a4paper,11pt]{article}
\usepackage{graphicx}
\usepackage{multirow}
\usepackage{amsmath}
\usepackage[a4paper,left=0.8in,right=0.8in,top=1.0in,bottom=1.0in]{geometry}
\usepackage[cmtip,arrow]{xy}
\usepackage{pb-diagram,pb-xy}
\usepackage{pst-plot, pst-coil, pstricks}
\usepackage{appendix}
 \usepackage{fancyhdr}
\usepackage{subfigure}
\usepackage{bm}
\usepackage{url}
\usepackage{cite}

\newcommand{\be}{\begin{equation}}
\newcommand{\ee}{\end{equation}}

\begin{document} 

\title{A comparative study of jet substructure taggers}

\author{
  Paloma Quiroga-Arias \\
  {\it LPTHE, UPMC Univ. Paris 6 and CNRS UMR7589, Paris, France}\\ \\
  Sebastian Sapeta \\
  {\it IPPP, Durham University, South Rd, Durham DH1 3LE, UK}
}
\date{}
 
\maketitle

\vspace{-19em}
\begin{flushright}
  IPPP/12/69\\
  DCPT/12/138\\
\end{flushright}
\vspace{14em}
  
\abstract{
We explicitly study how jet substructure taggers act on a set of signal and
background events. 
We focus on two-pronged hadronic decay of a boosted Z boson. The background
to this process comes from QCD jets with masses of the order of $m_Z$.
We find a way to compare various taggers within a single framework by
applying them to the most relevant splitting in a jet.
We develop a tool, {\tt TOY-TAG}, which allows one to get insight into what
happens when a particular tagger is applied to a set of signal or background
events. It also provides estimates for significance and purity.
We use our tool  to analyze differences between various taggers and potential
ways to improve the performance by combining several of them.

} 
  
\section{Introduction}\label{intro}

The substructure of boosted hadronically-decaying objects differs from that of
pure QCD jets whose mass falls in a compatible invariant mass window. Therefore
the usefulness of boosted jet substructure techniques for the identification of
hadronically decaying massive objects such as W/Z
bosons~\cite{Butterworth:2002tt}, Higgs
boson~\cite{Seymour:1993mx,Butterworth:2008iy,Plehn:2009rk, Kribs:2010hp,
Gallicchio:2010dq, Soper:2010xk, Richardson:2012bn} or the top
quark~\cite{Seymour:1993mx,Kaplan:2008ie,Ellis:2009su, Ellis:2009me} (for review
see \cite{Plehn:2011tg}).
The existing procedures, such as filtering~\cite{Butterworth:2008iy},
pruning~\cite{Ellis:2009me} or trimming~\cite{Krohn:2009th}, extract the
internal structure of jets by using information from the clustering procedure.
Those substructure algorithms are able to distinguish between heavy resonances
and pure QCD contributions, therefore allow for enhancing signal to
background ratio.

In the last years, there has been a proliferation of new taggers which are used
for the identification and cleaning of signal versus background (for the most
recent review see \cite{Altheimer:2012mn}). 
We have also already seen an impressive variety of experimental studies using
these techniques \cite{Abazov:2011vh, Chatrchyan:2013rla, ATLAS:2012am}.
Though obviously (subtle in some
cases) differences exist between various taggers and combining them may
lead to an enhanced statistical power~\cite{Soper:2010xk, Cui:2010km}, there
must be a significant amount of overlap between them as they are often based on
very similar principles.
Unfortunately, full algorithmic implementations of the taggers are complex and
if one further combines several taggers and optimizes them the whole procedure
becomes a black box with very limited possibility to gain an understanding of
what actually happens to the ensemble of the signal and background events.

In this paper we perform a study of a selection of the most commonly
used tagging methods in an attempt to understand and characterize where the
differences in the various definitions come from and to what substructure
features the different outcome after applying the taggers can be attributed. 
In order to make it possible, and to some extent analytically trackable, we
perform some approximations to the tagger definitions
and apply them to Monte Carlo simulations of a signal and the corresponding
background processes.  

From such simulations we take the hardest (fat) jet, which in the case of signal
corresponds to the hadronic decay of the boosted heavy resonance, and we analyze
its substructure and the performance of the various taggers.  For that purpose,
we developed an interactive tagging tool, \texttt{TOY-TAG}, which allows one to
apply different taggers to the desired signal and background files \emph{in a
user friendly way} 

The paper is organized as follows. In the next section we describe how we
simulate our signal and background events, which approximations we make to the
mass of jets (sec.~\ref{sec:m-splitt}) and we introduce the concept or relevant
splitting (sec.~\ref{sec:relsplitt}) that will enable us to analyze all the
taggers in the same framework. The tagger definitions and the approximations we
apply to them are given in sec.~\ref{sec:tagger-def}. In
sec.~\ref{sec:interactive-tagger} we introduce our {\tt TOY-TAG} tool an make
some comparisons to the exact implementations of the taggers. The tool is then
used in sec.~\ref{sec:results} to discuss differences between taggers and to
study how they perform in term of purity and significance, also in the case with
combinations of taggers. We summarize our work in sec.~\ref{sec:summary}.

\section{Study of taggers on Monte Carlo simulations}
\label{sec:MC}

Our aim is to explicitly study how the taggers act on a set of statistically
significant ensemble of signal and background events. 
We will focus on two-pronged hadronic decays of boosted electroweak bosons
reconstructed using fat jet substructure techniques.\footnote{A hadronically
decaying boosted heavy object is seen as a single (large radius, therefore
'fat') jet. Such jet is called a W-jet or Z-jet when the heavy object is an EW
boson.} The background to this process will be coming from QCD jets whose mass
is comparable with that of a signal jet.

\subsection{Event simulation and reconstruction}
\label{sec:event-sim-rec}

We have generated several million proton-proton collisions  at nominal LHC
energy $\sqrt{s}=$14 TeV for the processes Z+jet ($\to l^+l^-$+jet) and ZZ ($\to
l^+l^-$+hadrons) using Pythia 6.4~\cite{Sjostrand:2006za}. The event generation
was done at hadron level with the underlying event switched off. 
The latter can be justified by the fact that, as shall be explained in the
following subsection, our implementations of the taggers will analyze directly
the most relevant splitting in a jet. The original implementations, however,
iterate over many splittings until they get to the one which is the most
relevant. On the way to that splitting, they remove significant part of the
underlying event.  Since we access the most relevant splitting directly, we do
not have the mechanism of UE removal. Therefore, our implementations of the
taggers acting on the events without UE approximately correspond to the original
taggers acting on the events with UE.
We performed jet reconstruction using either the $k_t$~\cite{Catani:1993hr}
or the Cambridge/Aachen (C/A)~\cite{Dokshitzer:1997in,Wobisch:1998wt} sequential
recombination algorithm  with R=1 (so that both products of the decay of the
heavy object belong to the same jet), as implemented in FastJet
3.0.1~\cite{Cacciari:2005hq,Cacciari:2011ma,fastjet}. 
Different values of the cut for the jet transverse momentum $p_{T,\min}$ were
used between 400 and 800 GeV and the window for the mass of the hardest jet was
imposed at 70~$ < m_J < $~110~GeV. 
For both processes we look for the lepton pair
from the decay of the boson that reconstructs its mass, and in
the opposite hemisphere there is a fat jet which is kept if its mass falls
within the stated window. In the case of ZZ, such fat jet corresponds to our
signal, that is the hadronic decay of the boson, while in Z+jet constitutes the
background, with a QCD jet faking the mass of a Z boson, and therefore the
substructure of such jet is different from the genuine decay. 

\subsection{Mass of a jet and the relevant splitting}
\label{sec:m-splitt}

Amongst all the splittings that one finds in a signal jet, there is one which is
special since it does not come from a QCD branching but from a decay of a heavy
object. This splitting is responsible for most of the mass of a jet and the
substructure techniques are designed to search for it in order to separate
signal from background.  Therefore, we will refer to it as the \emph{relevant
splitting} throughout the paper. During the jet clustering procedure, the
relevant splitting corresponds to the merging of two subjets, which we will call
$J_1$ and $J_2$, into the jet $J$. The two main characteristics of this merging
(splitting) are
\begin{equation}\label{eq:zDelta}
z = \frac{p_{TJ_1}}{p_{TJ}}\,,\qquad
\Delta = \sqrt{(y_{J_1}-y_{J_2})^2+(\phi_{J_1}-\phi_{J_2})^2}\,,
\end{equation}
and the mass of the jet $J$ can be expressed in terms of these variables to
be approximately
\begin{equation}
\label{eq:jetmass}
m_J^2 \simeq z(1-z)p^2_{T,J}\Delta^2 + m_{J1}^2 + m_{J2}^2\,.
\end{equation}
Let us now briefly consider what approximations are made to get to the above
formula.  The exact mass of a jet, expressed in terms of its two parent subjets, is given by
\begin{equation}\label{eq:mjexact}
m_{J,{\rm exact}}^2 = 
2 m_{T,J1} m_{T,J2}\cosh(y_{J_2}-y_{J_1}) -
2 p_{T,J1} p_{T,J2}\cos(\phi_{J_2}-\phi_{J_1})
+m^2_{J1}+m^2_{J2} \, .
\end{equation}
When $m_{J1(J2)} \ll p_{T,J1(J2)}$, the mass of the jet can be approximated by
\begin{equation}\label{eq:mjapp1}
m_{J,{\rm app1}}^2 =  
2 p_{T,J1} p_{T,J2}\Big[\cosh(y_{J_2}-y_{J_1}) - 
\cos(\phi_{J_2}-\phi_{J_1}) \Big]
+m^2_{J1}+m^2_{J2} \, .
\end{equation}
This holds well if the two subjets have commensurate transverse momenta.
However, for asymmetric jets there is a significant correction to this
approximation. 
If we further assume that the radius of our jet $R$ is small we get
\begin{equation}\label{eq:mjapp2}
m_{J,{\rm app2}}^2= 
p_{T,J1} p_{T,J2} \Delta^2 
+m^2_{J1}+m^2_{J2}\, ,
\end{equation}
with $\Delta$ defined by Eq.~\eqref{eq:zDelta}. This works well for
sufficiently boosted jets.  For Z-jets a transverse momentum  of 400 GeV gives
validity to this approximation (see table~\ref{table:app} for the numbers as a
function of $p_T$).
The next step is to use the variable $z=p_{T,J1}/p_{TJ}$ and assuming $p_{T,J2}=(1-z)p_{TJ}$
\begin{equation}\label{eq:mjapp3}
m_{J,{\rm app3}}^2 = 
z(1-z)p_{TJ}^2 \Delta^2 +m^2_{J1}+m^2_{J2}\, .
\end{equation}
This is rather innocent as compared to the other approximations that we have
just discussed, especially in a boosted regime, where it does not introduce more
than a 3\% error.
One can make further approximation by dropping the masses of the subjets
\begin{equation}
\label{eq:mjapp4}
m_{J,{\rm app4}}^2 = 
z(1-z)p_{TJ}^2\Delta^2\, .
\end{equation}
This is equivalent to having a two massless particle jet, and it is a reasonable
assumption in the case of a signal jet, because, as it has been already stated,
its mass comes mainly from the relevant splitting. 

To estimate the error introduced by each level of approximation, we calculate
the percentage by which the mass is misestimated, averaging over all the events
considered, and the corresponding standard deviation 
\begin{equation}\label{eq:err-app}
\delta_i=\frac{m_J-m_{J,{\rm appi}}}{m_J} \, , \ \ \ \ \ \sigma_{\delta,i} \, .
\end{equation}
The results are summarized in table~\ref{table:app}.  Both the Z-jet and QCD
events were produced at $\sqrt{s} = 14\, \text{TeV}$ and selected with the
$[70,110]\, \text{GeV}$ mass window. No tagging was applied at this stage. We
see that the approximations corresponding to Eqs.~\eqref{eq:mjapp3}
or~\eqref{eq:mjapp4} lead to ${\cal O}(10\%)$ average effect on the mass of the
signal jets and ${\cal O}(20\%)$ for the background QCD jets (for the jets which
pass the above mass window selection cut).

Since we aim at performing the study of taggers in an analytic way,
approximations are necessary. After analysing the effects of different
degrees of approximation on the mass distributions, in what follows we will use
the results from Eqs.~\eqref{eq:mjapp3} or~\eqref{eq:mjapp4}.

\begin{table}[t]
\begin{center}
 \begin{tabular}{|c|c|c|c|c|c|c|c|c|}
 \hline
    & $\delta_1 (\%)$ & $\delta_2 (\%)$ & $\delta_3 (\%)$ & $\delta_4 (\%)$  & $\sigma_{\delta,1} $  & $\sigma_{\delta,2} $  & $\sigma_{\delta,3} $  & $\sigma_{\delta,4} $\\
  \hline
   Z-jet & & & & & & & & \\
\hline 
$p_T$=200 GeV & 5.0 & 4.9 & 7.3 & 12.9 & 2.9 & 3.4 & 3.6 & 5.7 \\
\hline
$p_T$=400 GeV & 5.2 & 5.1 & 5.9 & 11.9 & 3.2 & 3.3 & 3.5 & 6.0 \\ 
\hline
\hline
   QCD-jet & & & & & & & & \\
\hline 
$p_T$=200 GeV & 7.8 & 7.8 & 10.0 & 17.8 & 3.7 & 4.1 & 4.2 & 6.2 \\
\hline
$p_T$=400 GeV & 8.9 & 8.8  & 9.7 & 18.4 & 4.3 & 4.5 & 4.7 & 6.6 \\ 
\hline
 \end{tabular}
 \end{center}
  \caption{Amount by which the mass of a jet is underestimated when taking the
  different approximations from Eqs.~\eqref{eq:mjapp1}--\eqref{eq:mjapp4}.
  }
  \label{table:app}
 \end{table}

\subsection{Optimal determination of the relevant splitting}
\label{sec:relsplitt}

\begin{figure}[t]
\centering
\includegraphics[width=0.43\textwidth]{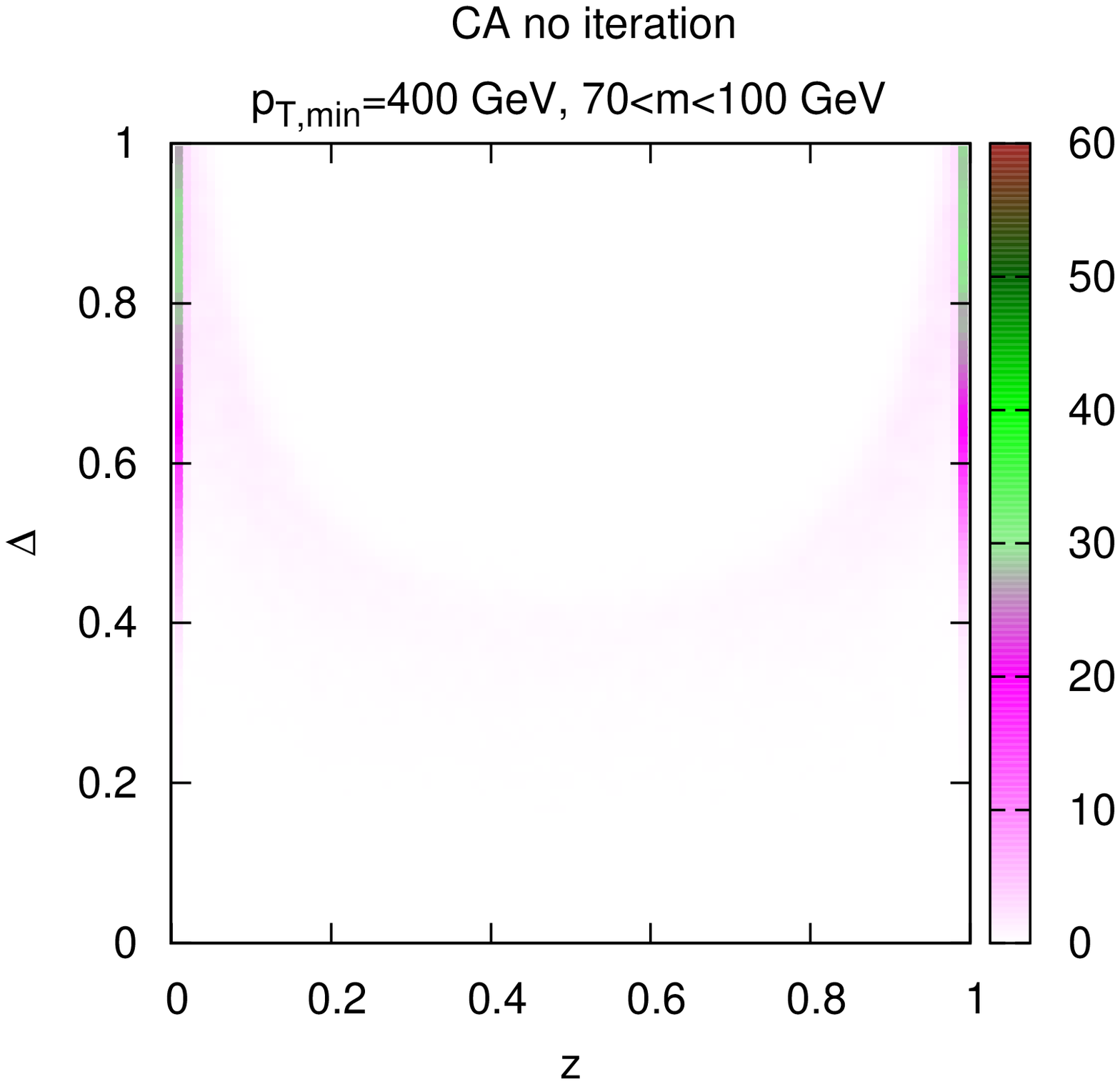}
\includegraphics[width=0.43\textwidth]{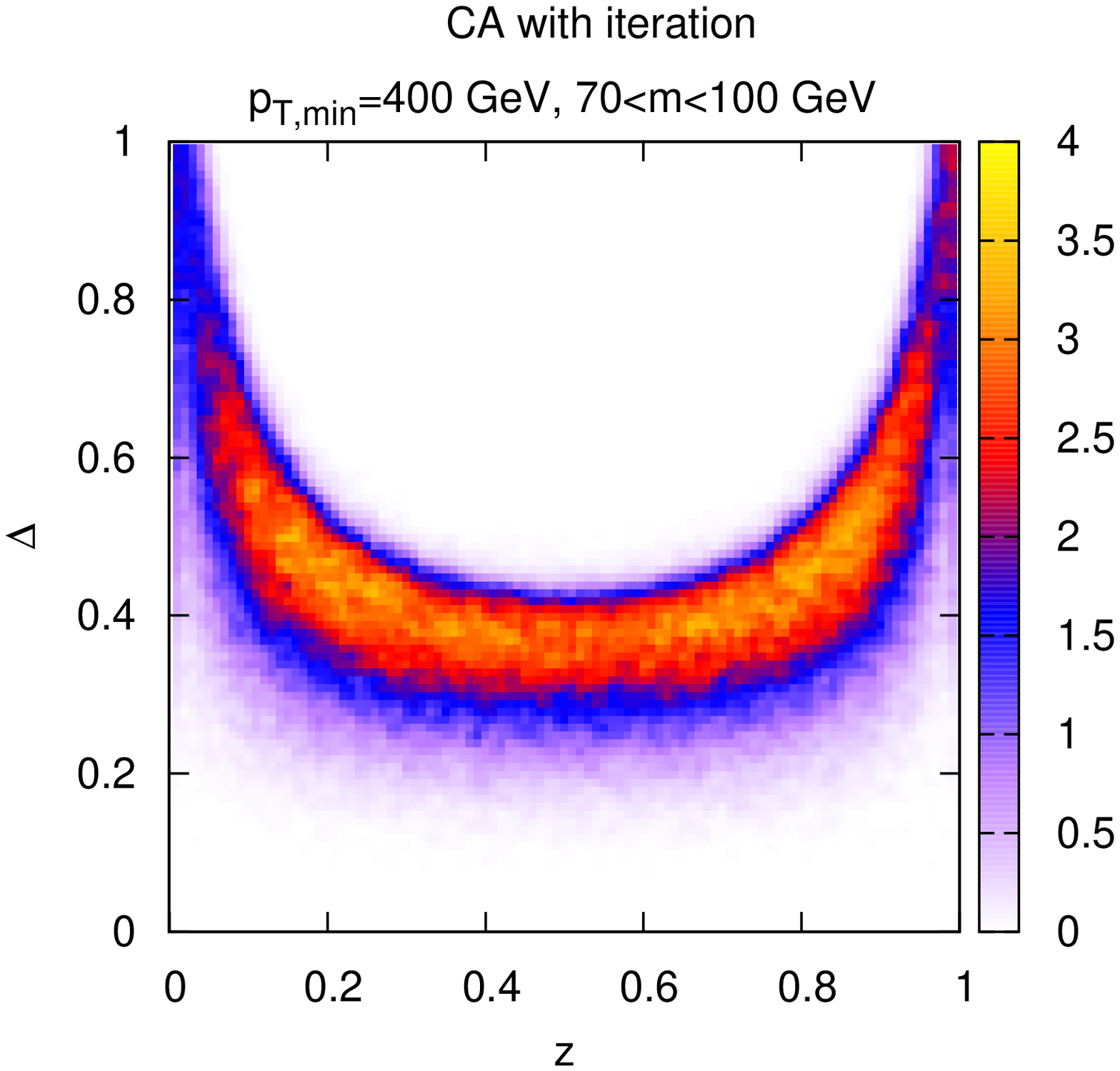}
\caption{\label{fig:CAit} Unclustering of the hardest jet with C/A algorithm.
Left: just one step in the unclustering process. Right: iteration in the
unclustering process until the relevant splitting is found (see text for more
detailed explanation). The vertical axis, whose values are represented by the
color gradient to the right of each plot, represent the cross section
$\frac{d \sigma}{d z d\Delta}$ in femtobarns. 
}
\label{fig:check-it-CA}
\end{figure}

\begin{figure}[t]
\centering
\includegraphics[width=0.43\textwidth]{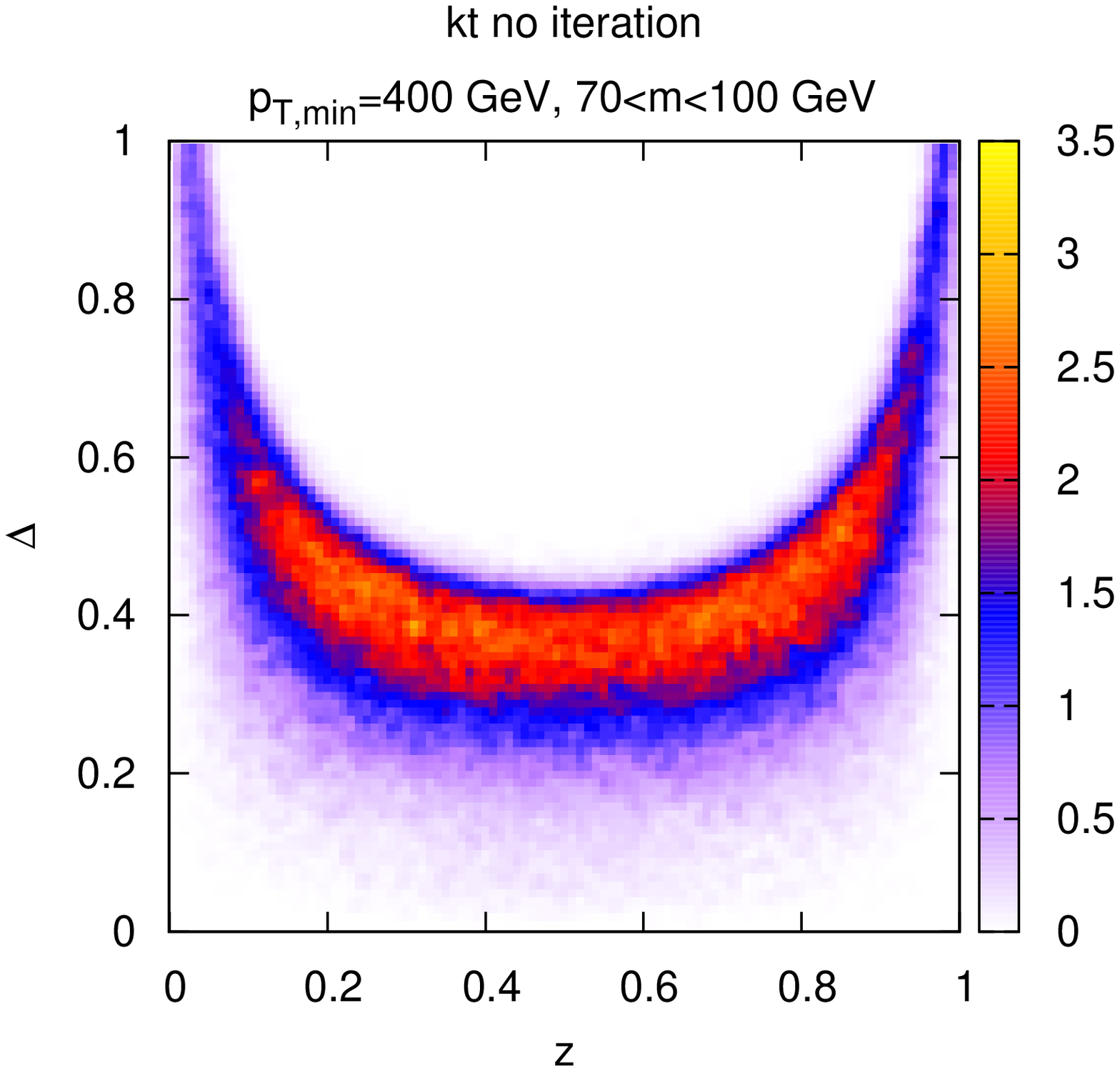}
\includegraphics[width=0.43\textwidth]{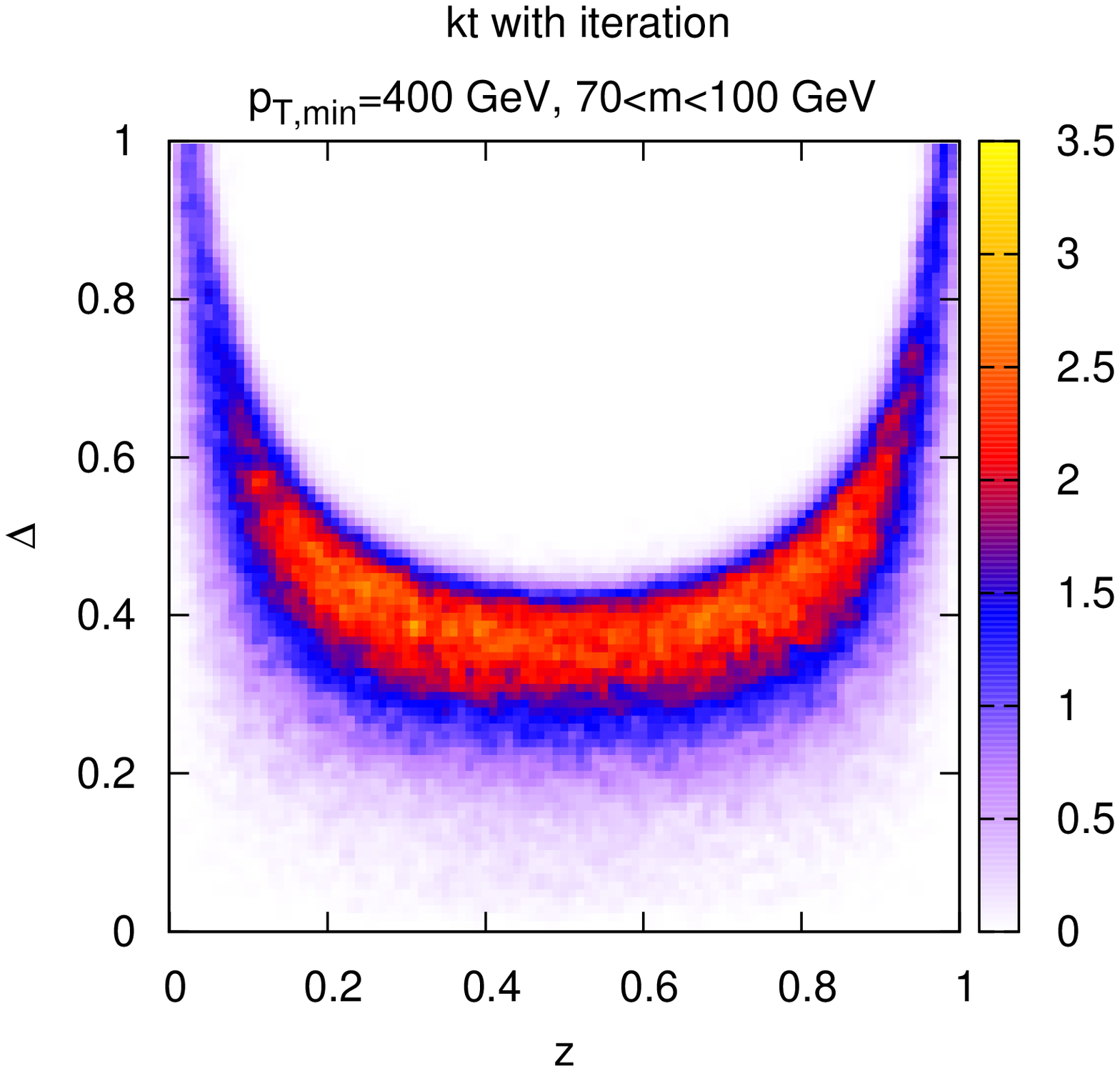}
\caption{\label{ktit} Same as in Fig.~\ref{fig:CAit} but for the $k_t$
algorithm.}
\label{fig:check-it-kt}
\end{figure}

In our analysis we will be interested in identifying the relevant splitting out
of the fat jet substructure in simulated events. Each event will be labelled
with the $(z,\Delta)$ pair corresponding to such splitting in order to study how
signal and background events differ from each other and how much of that
difference is captured by various tagging techniques.

Depending on which jet finding algorithm we take, the subjets coming from the
relevant splitting are merged at a different stage of its clustering.
Therefore, it is important to choose an algorithm which allows one to get to
this splitting as easily as possible and without biasing the sample.  Since we
are interested only in the infrared and collinear safe algorithms 
with well defined jet substructure, the choice is limited to either
C/A~\cite{Dokshitzer:1997in,Wobisch:1998wt} or $k_t$~\cite{Catani:1993hr} jet
algorithms.

The mass of a jet, Eq.~(\ref{eq:jetmass}), involves both the distance between
its two subjets, $\Delta$, and the ratio of the subjet transverse momentum to
the $p_T$ of the jet, $z =p_{T, \text{subjet}}/p_{T,\text{jet}}$. The subjet
used to compute this ratio is chosen randomly from the two available subjets.
Hence, sometimes it is the softer and sometimes the harder one. Therefore $z$
is in the range $[0,1]$.
For a fat jet resulting from the hadronic decay of a heavy object, both $z$ and
$\Delta$ are relatively large for the relevant splitting. Since the measure of the $k_t$ algorithm
involves both of these variables and is the largest when the two are maximal,
one expects that the relevant splitting will correspond to the last merging for
the $k_t$ algorithm.  This is not necessarily the case for the Cambridge/Aachen
algorithm since its measure involves nothing but the geometric
distance~$\Delta$.

To show that this is indeed the case, we performed the following study. We
generated a sample of boosted jets from the hadronic  decay of a Z boson with
the selection criteria described in sec.~\ref{sec:event-sim-rec}\footnote{The MC
generation of events is done with the underlying event turned off. We have
repeated the same exercise for ZZ with the underlying event as well for Zj with
Z decaying leptonically, each time finding the same behaviour as for the case of
ZZ with no UE described below.} and the cut on transverse momentum of the
hardest jet $p_{T,\min}=400\,\text{GeV}$ .
Next we proceeded to analyse the substructure of those jets, applying the same
procedure to the $k_t$ and C/A algorithms. We unclustered the last step of the
algorithm and used the two resulting subjets to determine $z$ and $\Delta$. Then
we dropped the lighter of the two subjets and repeated the above steps for the
heavier subjet. We continued this procedure each time checking before
unclustering whether our jet's mass was above 70~GeV. In the case in which the
mass dropped below 70~GeV we stooped iterating and stored the corresponding
$(z,\Delta)$ pair as a final value in the histogram.
The purpose of this analysis was to identify the relevant splitting, the main
idea being that if our unclustering happens at the relevant splitting the mass
of the heavier subjet should drop significantly below 70~GeV.

In Fig.~\ref{fig:check-it-CA} we show the corresponding results for the C/A
algorithm. The left histogram was obtained after just one unclustering so it
corresponds to the last merging in the clustering sequence regardless of whether
or not this merging is the relevant one. On the contrary, on the right hand
side, the values of $z$ and $\Delta$ correspond to the relevant splitting
determined with the iteration procedure described above. The two plots differ drastically
which tells us that the last merging in the C/A algorithm is hardly ever the
relevant one.  Most of the time it is just a large angle merging of highly
asymmetric objects.

This picture looks very different for the $k_t$ algorithm as shown in
Fig.~\ref{fig:check-it-kt}. Here, the histograms obtained after just one
unclustering and, alternatively, after the relevant splitting was identified,
are basically the same. This suggests that for the $k_t$ algorithm the last
splitting is most of the time the relevant one. In other words, the $(z,\Delta)$
pair of the two subjets from the first unclustering in the $k_t$ algorithm
should account for most of the mass of a jet. 

We remark that it was this feature of the $k_t$ algorithm that was exploited by
the seminal papers on jet substructure~\cite{Seymour:1993mx,Butterworth:2002tt}.
It also makes the $k_t$ algorithm particularly suitable for our further study of
the taggers since it allows one to get the $z$ and $\Delta$ of the relevant
splitting directly and obtain a tagger-independent profile of an ensemble of
events which then can be used to study how various taggers act on the signal and
background events.

\subsection{Taggers under study: definitions and approximations}
\label{sec:tagger-def}

As mentioned before, each event of the Monte Carlo simulation is labelled with
the (z, $\Delta$) pair of variables\footnote{It's $\Delta$/R, but we have always
used R=1.} corresponding to its relevant splitting and it is added to a 2D
colour map such as those shown in section~\ref{sec:relsplitt} (see
Fig.~\ref{fig:check-it-CA} and~\ref{fig:check-it-kt}). For this purpose, and
following the discussion from the previous section, we have used the
$k_t$-algorithm unclustering so that no iteration is needed. 
As we have already mentioned, the original formulations of the majority of the
taggers use the C/A algorithm, and therefore iterations are needed. By making
such choice the jets get cleaned in the process of finding the relevant
splitting throughout the iterations along the splitting process: in each
iteration the lightest of the two subjets is dropped, which is equivalent to
removing soft radiation. This cleaning process does not take place when using
$k_t$-unclustering. As we shall see in section~\ref{sec:validation}, where both
C/A with iterations and $k_t$ without iterations are used in MC simulations,
these two methods lead to similar results.  In order to study the performance of
different tagging techniques, we consider analytic definitions of the taggers
and take the necessary approximations, if any, so that they can be represented
within the same setup as the MC simulations, that is with each tagger shown as a
line which corresponds to a certain constant value of the tagging parameter.

The taggers considered in this study and the approximations made to
facilitate the analytic consideration and the consequent representation in the
(z,$\Delta$) plane together with the Monte Carlo simulations,  are presented
next. For each of the taggers we include the criteria applied in order to keep
signal and reject background events. Whenever the condition is satisfied, the
event is kept.

\begin{enumerate}

\item \textbf{Mass drop}

\noindent
As its name suggests, this tagger imposes a cut to the ratio of the mass of the
heavier of the subjets (J1) to the mass of the
jet~\cite{Plehn:2009rk,Butterworth:2008iy}
\begin{equation}\label{eq:mu_exact}
\frac{m_{J1}^2}{m_J^2} < \mu^2_\text{cut} \, ,
\end{equation}
to be accepted as a signal jet.  Therefore, in order to apply such tagger, it is
necessary to consider (at least) the heavier of the subjets as massive, so one
needs to go beyond the first approximation and take the mass of the jet to be
\begin{equation}\label{eq:mJapp_mu}
m_J^2=z(1-z)p_{TJ}^2\Delta^2+m_{J1}^2 \,,
\end{equation}
otherwise the tagger would be always trivially passed.
$m_{J1}$ in the above equation is the mass of the heaver subjet. In what
follows, we shall choose it to be constant and we shall fix it at some average
value that will be extracted from Monte Carlo.
Once normalized to the mass of the jet $m_{J}$, we have the following formula for the mass drop parameter as a function of the variables $(z,\Delta)$
\begin{equation}
\label{eq:mu}
1-\Delta^2z(1-z)\frac{p_{TJ}^2}{m_{J}^2} < \mu^2_\text{cut} \, .
\end{equation}
One thing that follows from the above formula and Eq.~(\ref{eq:mJapp_mu}) is
that there is a minimal value of $\mu^2_\text{cut}$ below which the mass drop
tagger (within our approximation) will have no effect on an event ensemble. This
minimal value corresponds to the point $\Delta=1$ and $z=\frac12$ and is given
by $\mu^2_\text{cut, min} = 1/(1+p_{TJ}^2/(4m_{J1}^2))$. 
This will act as a limitation of our current approximation of this
tagger and it will be less and less relevant for high values of $p_{TJ}$.

\item \textbf{Asymmetry cut}

\noindent
This tagger eliminates the most asymmetric configurations which generate big
masses but are most commonly coming from QCD radiation and not from the decay of
a heavy object. It is defined as~\cite{Butterworth:2008iy}
\begin{equation}
\label{eq:ycut}
\frac{p_{TJ}^2}{m_J^2}\rm{min}\left(z^2,(1-z)^2\right)\Delta^2> y_{cut}\, .
\end{equation}
In what follows, as in the case of mass drop, we shall use that
formula~(\ref{eq:mJapp_mu}) for the mass of the jet. That sets the upper limit
for the values of the tagger parameter which is $y_\text{cut, max} = 1/(1+4
m_{J1}^2/p_{TJ}^2)$. As we shall see later, this limit will have no practical
importance for the range of the jet transverse momenta of interest. We note
that, unlike in the case of the mass drop, the asymmetry cut is nontrivial even
in the massless subjet approximation to the jet mass from Eq.~(\ref{eq:mjapp4}).
However, as we shall see in sections~\ref{sec:interactive-tagger} and
\ref{sec:results}, by using finite $m_{J1}$, we reproduce important features of
the original implementation of this tagger.
 
\item \textbf{Jade distance normalized to $\mathbf{p_{TJ}}$}

The approximation to the jet mass from Eq.~(\ref{eq:mjapp4}) suggests the use of
the Jade distance~\cite{Butterworth:2007ke,Butterworth:2009qa} to construct the
tagger which rejects all jets whose two subjets are too close according to the
Jade measure. In order to get a dimensionless condition, we can normalize the
Jade distance to the transverse momentum of the jet. This leads to a criterion
which neither depends on the jet mass nor its $p_T$, but only on the ratio of
the transverse momentum of the subjet to the $p_T$ of the jet, $z$, and the
separation between subjets\footnote{It really is $\Delta$/R but, once again,
this study is restricted to R=1.}, $\Delta$,
\begin{equation}
\label{eq:modjade}
z(1-z)\Delta^2>J_{p_T,\text{cut}} \, .
\end{equation}
The viable range for the values of this tagger is $ J_{p_{T},\text{cut}}\in
[0,\frac14]$. We note that if we chose to normalize our Jade distance to the
mass of the jet we would have effectively reproduced the mass drop condition
from Eq.~(\ref{eq:mu}).

\item \textbf{Modified Jade distance normalized to $\mathbf{p_{TJ}}$}

\noindent
Similarly, a modified version of the Jade distance can be used to construct a
tagger, as was proposed in~\cite{Plehn:2009rk}.
If we normalize it to the transverse momentum of a jet we get

\begin{equation}
\label{eq:jadeprime}
z(1-z)\Delta^4>J_\text{mod, cut} \, .
\end{equation}
Similarly to the previous version of the Jade tagger, also here, the limits for
the values of the parameter are $J_\text{mod, cut} \in  [0,\frac14]$.

\item \textbf{1-subjettiness}

\noindent 
N-subjettiness~\cite{Thaler:2010tr,Stewart:2010tn} exploits the fact that the
pattern of the hadronic decay of a heavy object is reflected through the
presence of distinctive energy lobes corresponding to the decay products, as
opposed to QCD jets  which present a more uniformly spread energy configuration
(not aligned along the subjet axis). The inclusive jet shape N-subjettiness is
defined, in its generalized version as derived in~\cite{Thaler:2011gf}, as 
\begin{equation}\label{eq:nsub}
\tau_N=\frac{1}{d_0}\sum_k p_{T,k}\min\left((\Delta R_{1,k})^{\beta},...,(\Delta R_{N,k})^{\beta}\right) \, , \nonumber
\end{equation}
where $k$ runs over the constituent particles in the jet, $\Delta R_{J,k}=\sqrt{(\Delta y)^2+(\Delta \phi)^2}$, $\beta$ corresponds to an angular weighting exponent, and the normalization factor is
$d_0 = \sum_k p_{T,k}R^{\beta}$ with R being the jet radius.
The discriminating variable used when applying N-subjettiness in the
identification of two-prong hadronic objects is the ratio $\tau_2/\tau_1$, which
turns out to be smaller for signal than for background.
However, since in our simplified configuration we are considering the subjets
with no substructure, we can only consider the 1-subjettiness measurement,
$\tau_1$, without introducing additional variables that would not allow us to
represent this tagger in a ($z,\Delta$) map. In other words, we have a jet
($N=1$) with two constituent particles ($k=2$) whose distances to the jet axis
satisfy $\Delta R_{1,1}+\Delta R_{1,2}=\Delta$.

Therefore, we restrict ourselves to the 1-subjettiness case, which has the following functional form in terms of the characteristic variables $z$ and $\Delta$
\begin{equation}
\label{eq:1sub}
\Delta^{\beta}\left(z(1-z)^{\beta}+z^{\beta}(1-z)\right) > \tau_{1,\text{cut}} \, .
\end{equation}

The consideration of just $\tau_1$ for 2-prong hadronic objects, though not
optimal, is still meaningful, since the value of $\tau_1$ for a Z-jet will be
larger than that of a QCD-jet.
The viable range for the values of this tagger is $ \tau_{1,\text{cut}}\in
[0,\frac12]$.

\item \textbf{Pruning}

\noindent
Pruning~\cite{Ellis:2009su,Ellis:2009me} was designed to identify signal over
background configurations and at the same time to clean the former. By
definition it modifies the jet substructure in order to reduce the systematic
effects that obscure the reconstruction of hadronic heavy objects. It takes the
constituents of a jet putting them through a new clustering procedure in which
each of the branchings is requested to pass a set of cuts on kinematic
variables. If the cuts are not passed, then the recombination is vetoed and one
of the two branches is discarded. The conditions for each recombination $i,j\to p$ are
\begin{equation}
\frac{\min(p_{T,i},p_{T,j})}{p_{T,p}} < z_\text{cut} \ \  \rm{AND} \ \ \Delta R_{ij} > D_{cut} \, . \nonumber
\end{equation}
If both conditions are met, the merging does not take place and the softer
branch is discarded. One then continues the procedure with the remaining branch.
For the purpose of this study, similarly to what we have done for the other
taggers, we apply the pruning based criteria to the relevant splitting only. That
is, we require
\begin{equation}\label{eq:prun0}
\rm{min}\left(z,1-z\right)>z_\text{cut} \ \ \rm{OR} \ \  \Delta <D_{cut}\, ,
\end{equation}
for an event to be accepted. 
The dynamic condition on $\Delta$ is defined as
\begin{equation}\label{eq:Dcut}
D_\text{cut} = 2D\frac{m_J}{p_{TJ}} \, ,
\end{equation}
which results, if one takes the mass of the jet as given by Eq.~\eqref{eq:mJapp_mu}, in the following condition for pruning
\begin{equation}\label{eq:prun}
\min\left(z,1-z\right)>z_\text{cut} \ \ \text{OR} \ \  \Delta< \frac{m_{J1}}{p_{TJ}\sqrt{\frac{1}{4D^2}-z(1-z)}} \,,
\end{equation}
for the event to pass, with $m_{J1}$ being the mass of the most massive of the
two subjets. The parameters of the tagger can vary in the ranges $z_\text{cut}
\in [0,\frac12]$ and $D \in (0,p_{TJ}/(2m_J)]$. Note however, that for $D>1$ the
expression inside the square root in Eq.~(\ref{eq:prun}) will become zero and
then negative for some range around $z=\frac12$. This essentially means that
the condition for $\Delta$ is always satisfied in this region of $z$.
One needs to keep in mind that while this modified pruning still presents the signal to background discriminatory power, it is oblivious to its original capabilities to clean the signal jets that allowed for a more accurate reconstruction and an improved mass resolution.

\item \textbf{Trimming} 

This tagging procedure reclusters jet constituents into subjets with radius
$R_{\text{sub}}$ and then discards the contributions from subjets which do not
satisfy the condition
\begin{equation}\label{eq:trim}
p_{T,i} > f_{\text{cut}}\, \Lambda_{\text{hard}} \,,
\end{equation}
where $f_{\text{cut}}$ is a fixed dimensionless parameter and
$\Lambda_{\text{hard}}$ is some characteristic hard scale. 

In the case of our simplified configurations of two subjets at relevant
splitting, the trimming condition takes the form
\begin{equation}\label{eq:trim-approx}
\min(z,1-z) > \frac{f_{\text{cut}}\, \Lambda_{\text{hard}}}{p_{TJ}} \,,
\end{equation}
and assuming that for most of the jets $p_{TJ} \simeq p_{T,\min}$, the above
condition is nothing but a special case of pruning from Eq.~(\ref{eq:prun})
with $D_{\text{cut}}=D_{\text{cut},\min}$.

\end{enumerate}

The taggers from the above list were defined by various authors in contexts of
different studies. Admittedly however, the effect they have on a given ensemble
of events is not independent for each of them. It is difficult to quantify the
amount of overlap between different taggers in general.  However, it is possible
to find a common ground on which the taggers can be compared if some
approximations are applied to them. We believe that the approximations presented
in this section grasp the most essential features for each tagger.
 
Already at this point we can make several interesting observations.  First of
all, the asymmetry cut, as can be seen from Eqs.~\eqref{eq:mJapp_mu}
and~\eqref{eq:ycut}, involves not only the ratio of transverse momenta of the
subjets to the $p_T$ of the jet, $z$ or $1-z$, but also the distance between
them $\Delta$. It is therefore also rejecting those jets whose subjets at the
relevant splitting are geometrically too close to one another. In other words it
is doing part of the job of the group of taggers based on mass of the jet like
mass drop or Jade distances. This is important since the asymmetry cut and the
mass drop are often used together. 

It is also interesting to note in this context that the asymmetry cut kicks in
before the mass drop. This is because the former affects already the simplest
possible jet consisting of two massless partons in which case the condition from
Eq.~(\ref{eq:ycut}) takes the form $\min(\frac{z}{1-z},\frac{1-z}{z}) >
y_{\text{cut}}$. On the other hand, the same simplest configuration can never be
rejected by the mass drop tagger since the condition from
Eq.~(\ref{eq:mu_exact}) is always trivially satisfied for the case with
$m_{J1}=0$. Hence, for the mass drop to start having an effect, at least one of
the subjets has to have its own substructure which produces its non-zero mass.

By comparing Eqs.~(\ref{eq:prun}) and (\ref{eq:trim}) we can see that we should
expect some amount of overlap between pruning and trimming, at least in some
subregions of the parameter spaces of those taggers.

\section{Interactive tagger tool {\tt TOY-TAG}}
\label{sec:interactive-tagger}

We have developed an interactive tool, \texttt{TOY-TAG}, that allows the
user to apply the different taggers defined in section~\ref{sec:tagger-def},
separately or in combinations, to the desired signal and background files which
come in the form of a 2D surface plots for the $(z,\Delta)$ values of the
relevant splitting~(sec.~\ref{sec:relsplitt}). 
The script, available from \url{http://www.ippp.dur.ac.uk/~sapeta/toytag},
allows one to vary the values of the tagging parameters upon user request as
well as the minimum transverse momentum of the jet.
We have implemented all the taggers from section~\ref{sec:tagger-def} except for
trimming.  This is because, as discussed in sec.~\ref{sec:tagger-def}, to the
accuracy we are working in, the former is just a special case of pruning.

Since the information contained in the data files used by \texttt{TOY-TAG} is
limited to the values of $z$ and $\Delta$ for the relevant splitting, for those
taggers presenting an explicit dependence of the transverse momentum and mass of
the jet some minimal modelling is necessary. 
Since the cross section drops drastically with increasing transverse momentum
we set $p_{TJ}=p_{T,\min}$.
Concerning the mass, in order to bring the performance of the \texttt{TOY-TAG}
as close as possible to real tagging, in our implementation, the mass of the
more massive of the two subjets is taken into account as indicated by
Eq.~\eqref{eq:mJapp_mu}
\begin{equation}~\label{eq:tildemJapp}
m^2_{J} \simeq z(1-z)p_{TJ}^2\Delta^2 + m^2_{J1} \, , \nonumber
\end{equation}
with $m_{J1}$ being the mass of the heavier of the two parent subjets, which
will be set to a constant. In order to determine the value of $m_{J1}$,
we have studied the mass distribution of the heavier of the subjets in Monte
Carlo simulations (results not shown). 
The average mass of a subjet turns out to be larger for the background than for
the signal. The numerical values used in \texttt{TOY-TAG} are given in the
following subsection.

The \texttt{TOY-TAG} allows one not only to directly study which subgroups of
events are rejected by a particular tagger or a combination of taggers both for
signal and background jets, but also to get a first estimate of purity,  S/B,
and significance, S/$\sqrt{\rm{B}}$. Those quantities are computed by
integrating the 2D histograms in $(z,\Delta)$ above (or in the case of pruning
below) the line $\Delta = f(z)$, which is a representation of each tagger in
this plane. A concrete illustration will be given in sec.~\ref{sec:results}.

\subsection{Comparison to exact implementations of the taggers}
\label{sec:validation}

\begin{figure}[t]
  \centering
  \includegraphics[angle=-90,width=0.8\textwidth]{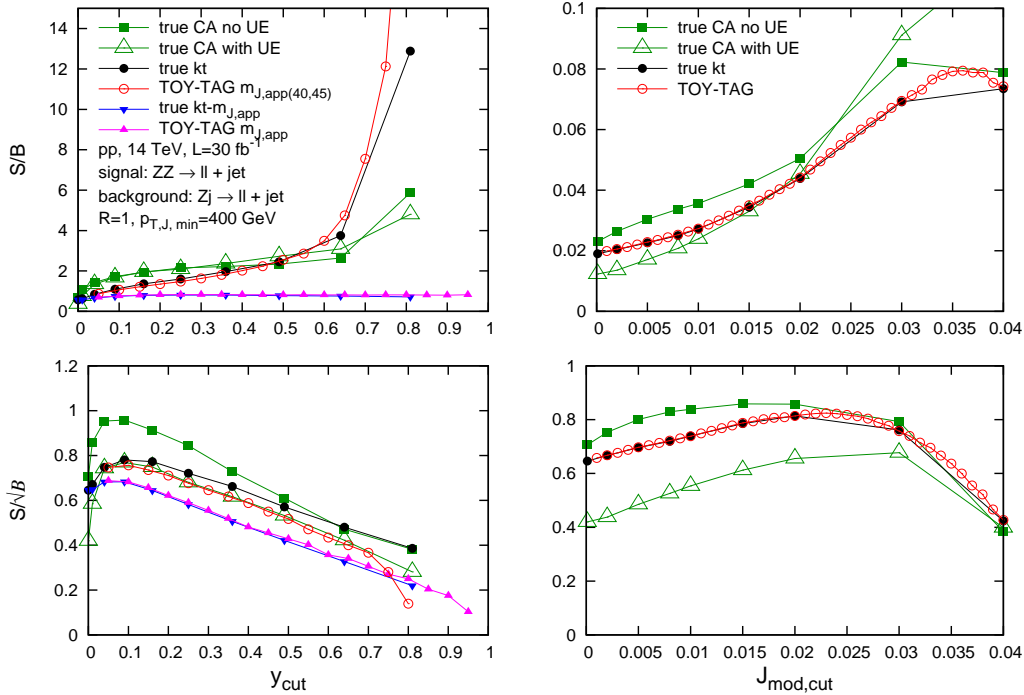}
  \caption{
  Comparison of \texttt{TOY-TAG} to exact taggers. S/B (upper row) and
  S/$\sqrt{{\rm B}}$ (lower row) for $\rm{y_{cut}}$, Eq.~\eqref{eq:ycut} (left
  panel), and $J_\text{mod, cut}$, Eq.~\eqref{eq:jadeprime} (right panel). 
  The minimum transverse momentum in the event sample used to produce this plot
  is $p_{T,\min}$=400 GeV. The label ``TOY-TAG $\rm{m_{J,app}}$'' refers to
  results obtained using the \texttt{TOY-TAG} with the approximation of
  Eq.~\eqref{eq:mjapp4} for the mass of the jet while the label
  ``TOY-TAG~$\rm{m_{J,app(40,45)}}$'' corresponds to the \texttt{TOY-TAG}
  using the mass approximation given by Eq.~\eqref{eq:mJapp_mu} with
  $m_{J1}$=40 and 45~GeV for signal and background respectively. 
  Results labelled as ``TOY-TAG'' correspond to the case in which there is
  no dependence on $m_J$ and $p_{TJ}$ and therefore no extra modelling is
  needed. The meaning of the rest of the labels is explained in the text.}
\label{fig:validation} 
\end{figure}
 
After all the simplifications that were necessary in order to construct a simple
tool which makes it possible to perform comparisons between various tagger, it
is interesting to check how realistic is the description 
provided by \texttt{TOY-TAG}.
We do this by computing the values of signal to background and signal to square
root of background for different taggers using \texttt{TOY-TAG} and
comparing them to the results from Monte Carlo simulations in which
the taggers were rigorously applied. The latter means using the exact
definitions of the taggers, unclustering with the C/A algorithm and using
iteration along the splitting chain until the tagging criteria is satisfied.

As argued in section ~\ref{sec:event-sim-rec}, \texttt{TOY-TAG} should be
applied to events without UE since it does not have the cleaning stage.  In
order to illustrate how big the effects of UE are and to give some estimate of a
potential uncertainty associated with it, we applied the exact taggers both to
events with and without UE, labeled as ``true CA with UE'' and ``true CA no UE''
respectively.

Also, in order to directly cross check our implementation and to quantify the
role played by the approximations adopted by \texttt{TOY-TAG}, we run the MC
simulation in which the taggers were applied to the first splitting using the
$k_t$-algorithm with no further iteration (see discussion in
section~\ref{sec:relsplitt}). 
Such results are labelled ``true kt''. On top of this, for those taggers
including explicitly a dependence on the jet mass, \texttt{TOY-TAG} needs
further approximation, as explained at the beginning of
sec.~\ref{sec:interactive-tagger}. In order to account for this extra
approximation, we run the MC simulation using the approximate value of $m_J$
which is used by \texttt{TOY-TAG}, Eq.~\eqref{eq:mJapp_mu}, and these results
are referred to as ``true kt-m$_{J,\text{app}}$''. 

In Fig.~\ref{fig:validation}, we show the results for signal to background and
signal to square root of background for two different taggers:\footnote{These
validation study has been performed for all the taggers included in
\texttt{TOY-TAG}. Though only two taggers are shown, the results found are
consistent for all of them, and the discussion in this section is applicable to
all of them.} one independent of the mass and transverse momentum of the jet,
modified Jade distance normalized to $p_T$, Eq.~\eqref{eq:modjade}, and another
one, the asymmetry cut, that has a dependence on $m_J$ and $p_{TJ}$ and hence
requires further modelling. 
The results shown in the figure correspond to the luminosity
$L=30\,\text{fb}^{-1}$. For \texttt{TOY-TAG} we consider different levels of
approximations explained before. 

We have found the values of the correction for the mass of the heaviest subjet,
Eq.~\eqref{eq:mJapp_mu}, by optimizing the behaviour of S/B and
S/$\sqrt{\rm{B}}$ as compared with tagging in MC. 
As mentioned earlier, we also studied exact masses of subjets from Monte Carlo
simulations. We can assert that the values for $m_{J1}$ found by optimizing 
Eq.~\eqref{eq:mJapp_mu} implemented in the \texttt{TOY-TAG} are
consistent with the actual masses of the subjets, being larger for a QCD-jet
than for a Z-jet. 
They are equal to 45~GeV and 40~GeV, for the background and the signal subjets
respectively.
Also, the fact that this values are weakly dependent on the
transverse momentum of the jet is in agreement with the results of the MC study.
We repeat that even though taking the constant value of $m_{J1}$ is a somewhat
crude approximation, it allows us to reproduce important features of the exact
versions of the taggers and at the same time enables us to gain some
understanding by being able to analytically study and compare the taggers in
their simplified forms. By introducing the parameter $m_{J1}$, which has the
meaning of the average mass of the hardest subjet, we mimic the real situation
with massive subjets and therefore the discussion based on our
implementations of the taggers becomes more realistic.

For taggers independent of $m_J$ and $p_{TJ}$, like the modified Jade distance
normalized to $\rm{p_{TJ}}$, shown in Fig.~\ref{fig:validation}~(right), the
performance of the \texttt{TOY-TAG} is very accurate up to large values of the
tagging parameter, in which the iteration process in the original C/A
formulation starts making a difference. 
\texttt{TOY-TAG} describes perfectly the results obtained in MC when
applying the tagger with $k_t$-unclustering.

In the case of taggers with explicit dependence on mass and $p_{TJ}$, the
situation is more complicated and a more careful interpretation of the results
is in place.  As shown in Fig.~\ref{fig:validation}~(left), we find perfect
agreement for the $y_\text{cut}$ tagger between the \texttt{TOY-TAG} and MC when
in both cases the mass of the jet is approximated by Eq.~\eqref{eq:mjapp4},
rejecting any mass contribution from the constituent subjets
(``$m_{J,\text{app}}$'' label in the plot). However, the results thus obtained
are far from those obtained from ``true kt '' and ``true CA'' on MC, especially
in the case of $S/B$. When the mass of the subjet is taken into account in
\texttt{TOY-TAG}, Eq.~\eqref{eq:mJapp_mu}, its performance improves
significantly, successfully describing the MC results from tagging with
$k_t$-unclustering (``true kt''). 
There are still some differences between improved \texttt{TOY-TAG} (and ``true
kt'') compared to the original tagger formulations with the C/A algorithm,
``true CA''. Especially S/B for the asymmetry cut tagger starts to be
very sensitive to the absence of iterations (in ``true kt'' and
\texttt{TOY-TAG}) for large value of $y_\text{cut}$. This is understandable
since by requiring high $y_\text{cut}$ we have less chances of finding a
sufficiently symmetric splitting in the background jet and that is why $S/B$
grows. The reason why the growth of S/B with $y_\text{cut}$ is smaller in the
case of C/A (hence for the case with iterations) is that by going deeper into
the substructure of a jet we increase the chance of finding a splitting which
satisfies the condition of the tagger. That will matter much more for the
background jet since its probability for the relevant splitting to be
symmetric is much smaller than in the case of a signal jet.
We can therefore conclude at this point that if the $y_\text{cut}$ tagger is
implemented with the $k_t$-algorithm (and therefore iterations play a minor
role), the \texttt{TOY-TAG} gives a very accurate results for $S/B$ and
$S/\sqrt{B}$. If however the C/A version of the asymmetry tagger is used, then
the result from \texttt{TOY-TAG} will be misestimated near the lower and the
upper limit of $y_\text{cut}$, still reproducing however the correct
qualitative behaviour.

We performed similar validation study for all the taggers implemented in
\texttt{TOY-TAG}. We have concluded that the agreement between the performance
of \texttt{TOY-TAG} and the true tagging applied to Monte Carlo simulations
stays within $40\%$ except for the values of the parameters near the limits
where only the qualitative behaviour of the taggers can be trusted.

In particular, the agreement for the significance, S/$\sqrt{\rm{B}}$, is quite
good for all taggers and all values of the tagging parameter, with the exception
of the mass drop tagger which for values of $\mu<$0.5, i.e. a very strict
tagging, shows larger deviations. This is expected since the mass drop is the
only tagger which requires, by definition, second order corrections (the fact
that the mass of the constituent subjets must be considered) and is directly
related to the value of $\mu^2_\text{cut, min}$ introduced after Eq.
(\ref{eq:mu}).  The results of \texttt{TOY-TAG}  not being reliable to high
accuracy for small values of mass drop does not have such dramatic consequences
since the typical values used for this tagger in studies involving jet
substructure are above 0.6~\cite{Butterworth:2008iy}. 

The quality of the description of $S/B$ by \texttt{TOY-TAG} is slightly worse
and not so generic. It is different for different taggers and it depends on the
values of the parameters. In particular, for taggers independent of the mass and
transverse momentum of the jet, Eqs.~\eqref{eq:modjade}-\eqref{eq:prun}, the
agreement between \texttt{TOY-TAG} and tagging in MC is remarkable when the
unclustering is performed with the $k_t$-algorithm, and remains within 20\% with
respect to true C/A tagging, provided that no extreme values of the tagging
parameters are used (for instance, for $J_{p_T,\text{cut}}>0.03$, the
reliability of the interactive tagger tool is deficient). On the other hand, the
agreement between \texttt{TOY-TAG} and true tagging on MC deteriorates for
taggers which depend on mass and $p_T$ of the jet. As shown before, for the
$y_\text{cut}$ tagger, the description by the \texttt{TOY-TAG} remains within
$\sim$ 30-40\% for moderate values of the tagger and a similar statement can be
applied to mass drop. The user should be aware of such limitations (at high and
low values of the parameters) at all times. 

The real jet substructure is obviously more complex than what can be inferred
form the relevant splitting. We would like to emphasize that our objective was
not to construct a tool which would replace dedicated jet substructure analysis
but rather to deliver a simple program which is able to provide some insight
into what happens when a particular tagger is applied to an ensemble of signal
or background events. At the same time, as discussed in this section, it is
capable of providing semi-quantitative (and often even fairly accurate) results
for $S/B$ and $S/\sqrt{B}$. Dedicated analysis cannot be replaced if one wants
to properly optimize the taggers but they often resemble black boxes and a lot
of information about how the taggers act on a particular event set is
inaccessible.  This information can be obtained with our \texttt{TOY-TAG}, which
in other words means that the tool that we provide is complementary to the above
mentioned dedicated substructure analyses.

\section{Studying taggers with {\tt TOY-TAG}}
\label{sec:results}

The results presented in this section are extracted from  the interactive tagger
tool \texttt{TOY-TAG} introduced and validated in the previous section.
We use \texttt{TOY-TAG} to illustrate and discus features of the taggers. All
results shown in this section were obtained with the
$p_{T,\min}=400\,\text{GeV}$ cut on jets, although similar conclusions are
valid for higher values of $p_{T,\min}$.

\begin{figure}[t] 
  \centering
  \includegraphics[width=1.0\textwidth]{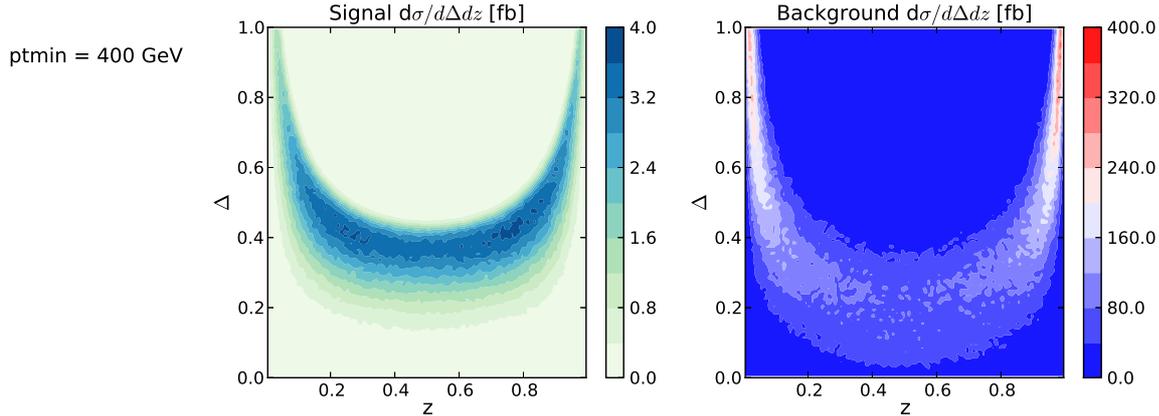}
  \caption{
  Histograms of $(z,\Delta)$ of the relevant splitting for signal and
  background events with the criteria from sec.~\ref{sec:event-sim-rec}.
  The fat jet is found with the $k_t$ algorithm with R=1.
  }
  \label{fig:sig_bkg_colormaps}
\end{figure}

Fig.~\ref{fig:sig_bkg_colormaps} shows the $(z,\Delta)$ histograms for the
relevant splitting of the signal and background jets obtained with the selection
criteria from section~\ref{sec:event-sim-rec}\footnote{In particular the U-type
shape of the histogram results from the window cut on the mass of the jet.}. 
It illustrates the main difference in the substructure between the Z-jets and
the QCD jets. For the case of hadronically decaying Z, most of the events gather
near $z\sim\frac12$ whereas the QCD events are concentrated close to
$z\sim 0$ and $z \sim 1$.
This is the difference that essentially all taggers try to exploit in order to
keep the most of the signal events and reject the most of the background events.
Another difference between the signal and the background events is that the
latter are much more abundant which can be seen by comparing the color scale in
the histograms in Fig.~\ref{fig:sig_bkg_colormaps}.

Since the majority of the background events concentrates at small momentum
fraction and large angles, those taggers which are able to reject that region of
the $(z,\Delta)$ plane will be preferred. Also, signal events are mostly
symmetric, hence removing the mentioned portion of the plane will not decrease
the signal too much. In the region of more symmetric $z$, the background events
have smaller values of $\Delta$ than the signal events.
We note in particular that some of the background events have very small
$\Delta$ which, based on Eq.~(\ref{eq:mjapp4}), could naively suggest that the
corresponding jet mass is below the cut used in the selection. We checked that
those jets are massive enough to pass the selection because their subjets have
finite masses which should not be neglected. In other words, the massless subjet
formula~(\ref{eq:mjapp4}) is too crude an approximation for the background jets
and one obtains much more realistic results with jet mass computed from
Eq.~(\ref{eq:mJapp_mu}). 
This is precisely what we use, for those taggers which depend explicitly on
$m_J$, in our implementations in {\tt TOY-TAG}, as discussed in
section~\ref{sec:tagger-def} and at the beginning of
section~\ref{sec:interactive-tagger}.
Putting the above things together, we can conclude that those taggers for which
the functional form of $\Delta (z)$ present a sharp slope at low $z$ and $1-z$,
and a rather flat behavior in the intermediate region, will provide best
cleaning of the signal.

\subsection{The taggers: differences and similarities}

\begin{figure}[t] 
  \centering
  \includegraphics[width=1.0\textwidth]{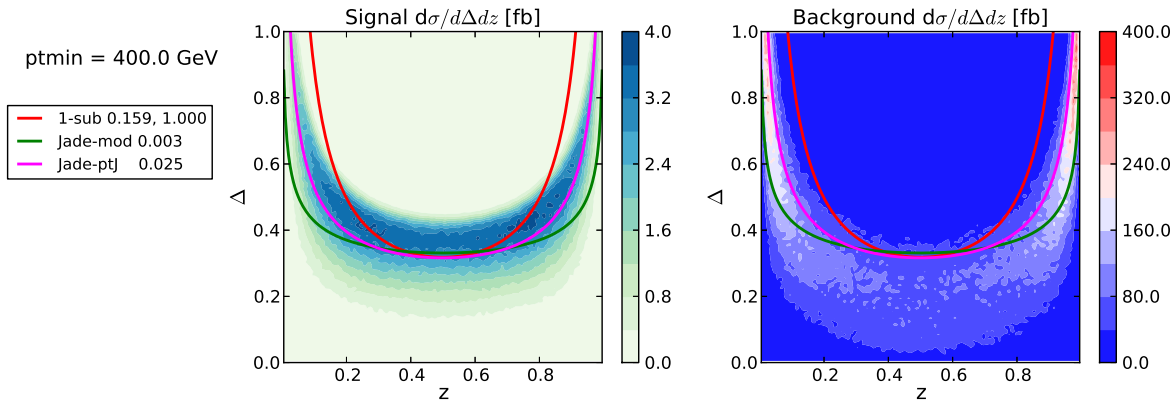}
  \caption{
  Comparison of taggers independent (within our approximations) of $p_{TJ}$ and
  $m_{J}$. The fat jet is found with the $k_t$ algorithm with R=1. In the case of 1-subjettiness the angular weighting exponent was set to $\beta=1$.
  }
  \label{fig:comp-Jades}
\end{figure}

Fig.~\ref{fig:comp-Jades} illustrates the first group of taggers, namely those
which do not depend on the jet mass and jet transverse momentum. These are:
1-subjettiness, Jade distance normalized to $p_{TJ}$ and the modified Jade
distance.
We remind the reader that  for each curve, all events above it, that is those
having larger $\Delta$ for a given $z$, are kept by the tagger and all events
below are rejected. This is true for all the taggers discussed in this paper
except for pruning as will be explained later.
As can be seen from Eqs.~(\ref{eq:modjade}), (\ref{eq:jadeprime}) and
(\ref{eq:1sub}), the main difference in the definition of those taggers is in
the power of $\Delta$.
Fig.~\ref{fig:comp-Jades} is an example showing what happens if we decide that
we want to keep the same level of symmetric events with $z\sim\frac12$ with each
of these three taggers.
We see that is such a case the events with low $z$ or $1-z$ are the most
effectively removed the lower the $\Delta$ power in the tagger definition. That
is why 1-subjettiness leaves almost no background events whereas modified Jade
accepts most of them. As we see in Fig.~\ref{fig:comp-Jades}, the situation is
more complex however since 1-subjettiness tends to remove also a significant
part of the signal events that we would like to keep, so it may turn out that
it is more preferable to use Jade-pTj which, due to its shape accepts much of
the signal events. This dilemma can be solved by computing the ratios of
signal/background and signal/$\sqrt{\text{background}}$ which we shall do in the
following subsection.

\begin{figure}[t] 
  \centering
  \includegraphics[width=1.0\textwidth]{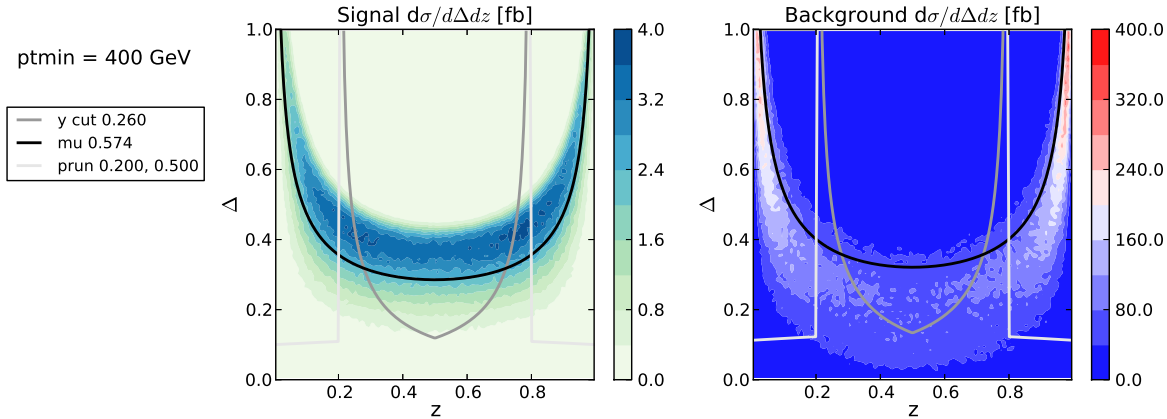}
  \caption{
  Comparison of three distinctly different taggers: mass drop, asymmetry cut and
  pruning. The fat jet is found with the $k_t$ algorithm with R=1.
  }
  \label{fig:md-ycut-pun}
\end{figure}

Fig.~\ref{fig:md-ycut-pun} shows three distinctly different taggers: mass drop,
asymmetry cut and pruning. Each of them is taken at an example value of the
parameter(s) for illustration. All events above the curve, for mass and ycut,
and below the pruning curve are accepted.
We see that the mass drop follows closely the shape of the color maps but in
such a way that most of the signal events are above the curve and most of the
background events below. On the contrary, the asymmetry cut and the pruning have
very different shapes than the histograms. These taggers are focused on removing
the low $z$ and low $1-z$ events and this is the main feature that they have in
common. We see from Fig.~\ref{fig:md-ycut-pun} that they could even mimic one
another in certain subregions of the parameter space. The main difference
between pruning and asymmetry cut, as illustrated in the figure, is that the
latter also removes part of the background events which are symmetric in $z$.
This feature comes from the fact that, as already mention in
sec.~\ref{sec:tagger-def}, the asymmetry cut tagger, even in its simplified
version from Eq.~(\ref{eq:ycut}), entangles $\Delta$ and $z$ whereas pruning
cuts on each of them separately.
It is interesting to note that the mass drop tagger and the asymmetry tagger or
pruning are complementary and one should expect to improve performance by
combining them as it was indeed done in a number of
analyses~\cite{Butterworth:2008iy, Cui:2010km}.

\subsection{Purity, significance and combinations of taggers}

One of the useful features of {\tt TOY-TAG} is that it can compute the purity of
the signal, $S/B$, and the significance, $S/\sqrt{B}$, for the simulated events
and the modelled taggers. This allows one to get a first estimate of the
performance of the taggers as a function of theirs parameters and, most
importantly, to study potential gains to be had by combining different taggers.

In Fig.~\ref{fig:comp-Jades} we had three different taggers with the parameters
chosen such that they accept similar number of events near $z\sim\frac12$. The
corresponding purity and significance values $(S/B, S/\sqrt{B})$ are: 
$(0.091, 0.71)$ for 1-subjettiness,  $(0.057, 0.82)$ 
for Jade-pTj and $(0.03, 0.65)$ for
modified Jade. 
As expected from the figure, we see that the modified Jade performs the
worst simply because it accepts too many background events with low $z$ or low
$1-z$. On the contrary  ``1-sub'' and ``Jade-pTj'' reject most of those events
which results in much better performance in terms of $S/B$ and $S/\sqrt{B}$. It
is also interesting to note that even though 1-subjettiness rejects more of the
signal events near $z\sim\frac12$ compare to the ``Jade-pTj'' it still does
better than the latter due to the fact that it also removes more of the
background events.

\begin{figure}[t] 
\includegraphics[width=0.3\textwidth, angle=-90]{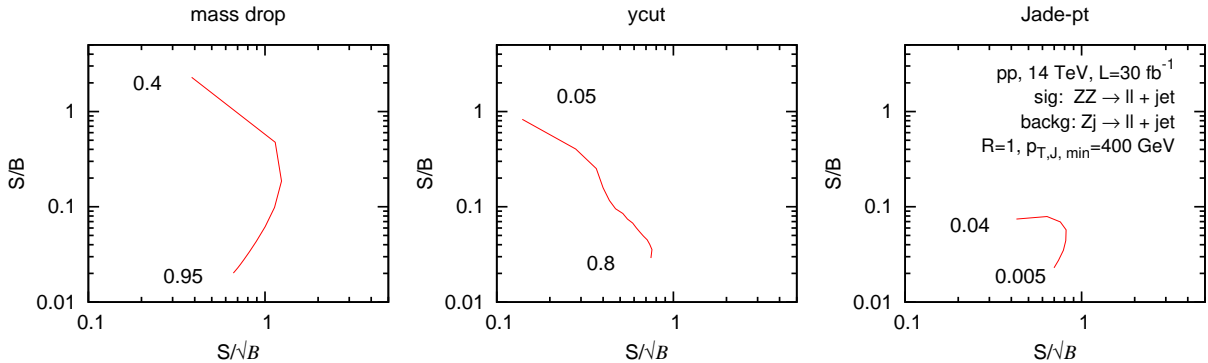}
  \caption{
  S/B vs. S/$\sqrt{\rm{B}}$ for different taggers applied to jets with $p_T >400
  \text{GeV}$. The numbers in the plots correspond to the values of tagger
  parameter at the extremes of each curve.}
  \label{fig:SBvsSsqrtB}
\end{figure}

{\tt TOY-TAG} allows for a scan over the full range of the tagger parameter,
which makes it easy to study $S/B$ vs. $S/\sqrt{B}$ relation. Some examples of
such plots are shown in Fig.~\ref{fig:SBvsSsqrtB}. 
Even though these results should be regarded as semi-quantitative, they provide
useful information about the expected trends as well as first estimates which
can help identifying potentially interesting ranges of tagger parameter.

\begin{figure}[t] 
  \centering
  \includegraphics[width=1.0\textwidth]{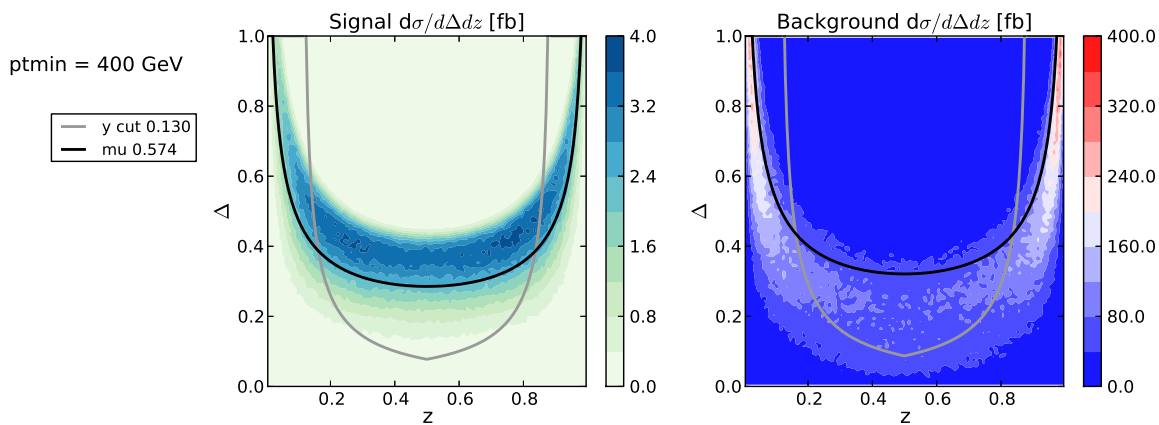}
  \caption{Combination of mass drop and asymmetry cut. The fat jet is found with
  the $k_t$ algorithm with R=1.
  }
  \label{fig:md-ycut-comb}
\end{figure}

Another interesting question that can be studied with {\tt TOY-TAG} is what
happens to purity and significance when two taggers are combined. An
illustration is given in Fig.~\ref{fig:md-ycut-comb} for the mass drop and
asymmetry cut.  It is clear from that figure that the asymmetry cut alone does
very well for the events with low $z$ and $1-z$ by rejecting a lot of background
in that region.  However, ycut alone, still keeps many relatively symmetric
background events with low $\Delta$. Those events could be rejected by imposing
higher values of the ycut parameter but that would lead to a drastic drop of the
signal events as well.
However, if the mass drop is added on the top of the asymmetry tagger, most of
these low $\Delta$ background events are rejected while most of the
corresponding signal events in that region are kept. That allows one to improve
the performance. In terms of the $(S/B,S/\sqrt{B})$ pair we have $(0.039,0.75)$
for ycut alone, $(0.073,1.1)$, for mass drop alone and $(0.11,1.1)$ for the
combination. Hence, we see that by combining the two tagger we were able to
improve the purity of our signal while keeping the same significance.

\section{Summary}
\label{sec:summary}

The difference between substructure of boosted jets coming from decays of heavy
objects and that of the QCD jets is widely used as a powerful tool to
discriminate between signal and background in searches for heavy objects.
The real substructure of jets is complex and full algorithmic implementations of
taggers often turn them into a black box, seriously limiting the chance of
understanding of what actually happens to the ensemble of the signal and
background events. This is inevitable if one aims at optimizing an analysis for
the highest performance. However, it is also interesting to try to obtain some
insight into what happens when a particular tagger or a group of taggers is
applied.

In this paper, we explicitly studied how jet substructure taggers act on a set
of signal and background events.  We considered ZZ ($\to l^+l^-$+jet) as signal
and Z+jet ($\to l^+l^-$+jet) as background. In order to be able to compare all
the taggers within a single framework we studied only how they act on the most
relevant splitting in a jet, that is the one which is responsible for most of
the jet mass. We have checked that this approximation grasps the most important
features and it allows for a semi-quantitative analysis.

We developed an interactive tool, \texttt{TOY-TAG}, that allows one to apply
approximate versions of the different taggers, or their combinations, to the
signal and background files and to study what happens when the tagger parameters
are varied.  It also provides estimates for the ratios of signal/background and
signal/$\sqrt{\text{background}}$. 
For some taggers, we introduced extra modeling, which involved mass of the
heavier subjet, and which allowed us to reproduced important features of the
exact versions of the taggers.
We view \texttt{TOY-TAG} as being complementary to dedicated substructure
analyses as it can provide some guidance and understating of the way a
particular tagger acts on an set of signal and background events.

We used our tool to discuss main differences between taggers. We
illustrated how various definitions of taggers translate into the regions of
$(z,\Delta)$ of the relevant splitting that are accepted or rejected. 
We identified the areas where different taggers overlap. On the other side, we
emphasized which distinct differences between the taggers lead to an improvement
of performance when several taggers are combined.

We believe that TOY-TAG, in its current form,  can serve many more analyses.
We also plan to extend it such that it is able to handle broader class of
input signal and background files as well as to apply any tagger defined as a
function $\Delta(z)$.
The  {\tt TOY-TAG} tool is available from the website
\url{http://www.ippp.dur.ac.uk/~sapeta/toytag}.

\section*{Acknowledgements}
We are grateful to G. Soyez for useful discussions throughout the development of
this work. We thank M. Cacciari and G. Salam for their valuable comments on the
manuscript and the {\tt TOY-TAG} tool. We thank J. Thaler for pointing out the generalized formulation of N-subjettiness. The work of PQA is supported by the
French ANR under contract ANR-09-BLAN-0060.

\bibliography{taggers-paper}
\bibliographystyle{unsrt} 

\end{document}